\documentstyle[prb,aps,epsf,multicol]{revtex}
\begin{document}

\title{\Large{\bf Ferrohydrodynamics: testing a new magnetization equation}}
\author{Mark I. Shliomis}
\address{Department of Mechanical Engineering, Ben-Gurion University of the Negev,\\
P.O.B. 653, Beer-Sheva 84105, Israel}
\date{\today}
\maketitle

% For wide abstract
\draft
\tighten

\begin{abstract}
A new magnetization equation recently derived from irreversible
thermodynamics is employed to the calculation of an increase of
ferrofluid viscosity in a magnetic field. Results of the
calculations are compared with those obtained on the basis of two
well-known magnetization equations. One of the
two was obtained phenomenologically, another one was derived microscopically
from the Fokker-Planck equation. It is shown that the new
magnetization equation yields a quite satisfactory description of
magnetiviscosity in the entire region of magnetic field strength
and the flow vorticity. This equation turns out to be
valid -- like the microscopically derived equation but unlike the
former phenomenological equation -- even far from equilibrium, and
so it should be recommended for further applications.

\end{abstract}

% For wide abstract
\begin{multicols}{2}
\narrowtext

The conventional set of ferrohydrodynamic equations [1-3] consists
of the equation of ferrofluid motion, the Maxwell equations, and
the magnetization equation. The latter was first derived thirty
years ago by the author [1]:
\begin{equation}
\frac{d\,{\bf M}}{d\,t}={\bf\Omega\!\times\! M-}\frac{1}{\tau}\,%
({\bf M-M}_{0})-\frac{1}{6\eta \phi }\,{\bf M\!\times\!%
(M\!\times\! H)} \eqnum{1}
\end{equation}
(${\rm Sh\,'72}$). Here ${\bf M}$ stands for the ferrofluid magnetization 
under the magnetic field ${\bf H}$ and the flow vorticity $2{\bf \Omega
=\nabla\times v} ,\;\eta$ is the fluid viscosity, $\phi=nV$ the
volume fraction of magnetic grains in the liquid, $n$ is their number 
density, $V$ the volume of a single particle, and $\tau=3\eta V/k_BT $ 
is the Brownian time of rotational particle diffusion. When the fluid is 
at rest in a stationary magnetic field, its {\em equilibrium} magnetization
${\bf M}_0$ is described by the Langevin formula
\begin{equation}
{\bf M}_{0}=nmL(\xi )\frac{\bbox{\xi}}{\xi},\;\;\; {\bbox{\xi}}%
=\frac{m{\bf H}}{k_{B}T},\;\;\; L(\xi )=\coth \xi -\xi ^{-1},
\eqnum{2}
\end{equation}
where $m$ is the magnetic moment of a single particle. The
phenomenological equation (1) generalizes the Debye relaxation
equation in case of {\sl spinning} magnetic grains. The spin originates
from viscous and magnetic torques acting upon the particles -- see
below Eq. (6).

Soon after [1], Martsenyuk, Raikher and Shliomis [4] (MRSh) proposed
another magnetization equation derived microscopically from the
Fokker--Planck equation. They have employed for the purpose an
original {\sl effective field method} (EFM) described in detail in [5]. 
According to the method, an arbitrary nonequilibrium magnetization $\bf M$ 
is considered at any moment as an equilibrium one in a certain -- 
specially prepared -- effective field ${\bf H}_{\rm e}$, that is
\begin{equation}
{\bf M}=nmL(\zeta)\,{\bbox{\zeta}}/{\zeta}\,,\;\;\;\;\;
{\bbox{\zeta}}=m{\bf H}_{\rm e}/k_{B}T\,. \eqnum{3}
\end{equation}
The magnetization (3) relaxes to its equilibrium value (2) as the
effective field ${\bf H}_{\rm e}$ (or ${\bbox{\zeta}}$) approaches
the true field ${\bf H}$ (or ${\bbox{\xi}}$). This relaxation
process is governed by the equation [4,6]
\begin{eqnarray}
\frac{d\,{\bf M}}{d\,t}={\bf\Omega\!\times\! M}-\frac{\left[\zeta ^2-%
({\bbox{\xi\zeta}})\right]}{\tau \zeta ^2}\,{\bf M}\nonumber\\
-\frac{\left[\zeta-L(\zeta)\right]}{6\eta \phi \,\zeta
L^2(\zeta)}\,{\bf M\!\times\!(M\!\times\! H)}.\;\;\;\;\; ({\rm MRSh})
\eqnum{4}
\end{eqnarray}
Equations (3) and (4) determine the dependence ${\bf M}(t;\, {\bf
H, \Omega})$ in an implicit form, dimensionless effective field
${\bbox{\zeta}}$ being the parameter.

It is well-established that Eq. (4) describes very fine real
ferrofluids. Comparison with computer simulation of the Brownian
motion in orientational space [7,8] indicated that the EFM yields
quite accurate results for any values of $\xi$ and $\Omega\tau$.
The same conclusion has been made in [6] under comparing of
solutions of Eq. (4) with the results of numerical integration of
the Fokker--Planck equation. At that time, the calculations [6-8]
have shown that the phenomenological equation (1) is valid for any
field magnitudes but only sufficiently small vorticities,
$\Omega\tau<1$. The applicability of (1) in the case
$\Omega\tau\ll 1$ was corroborated by numerical [9] and analytical
[10] solutions of the Fokker--Planck equation as well. Therefore,
under consideration of a weakly nonequilibrium situation, one
should give preference to Eq. (1) as it is far simpler for
analysis than (4). The latter, however, guarantees the correct
description of magnetization even if its deviation from
equilibrium value (2) is large, $\Omega\tau\gg 1$, that is when
Eq. (1) leads to erroneous results.

Let us show that a new phenomenological magnetization equation
derived recently [11] from irreversible thermodynamics,
\begin{equation}
\frac{d\,{\bf H}_{\rm e}}{d\,t}={\bf \Omega\!\times\! H}_{\rm e}-%
\frac{1}{\tau}\,({\bf H}_{\rm e}-{\bf H})-\frac{1}{6\eta \phi} %
\,{\bf H}_{\rm e}\!\times\!({\bf M}\!\times\!{\bf H}) \eqnum{5}
\end{equation}
(${\rm Sh\,'01}$), is free from the above mentioned shortcoming. Namely, 
being simple enough, the equation is valid even far from equilibrium. 
As a checking on applicability of the old (1) and the new (5)
phenomenological equations we have chosen their predictions about
the {\sl rotational} or {\sl spin viscosity} $\eta_r$ of
ferrofluids. Below we shall compare $\eta_r$ obtained from (1) and
(5) with its almost exact value resulting from the EFM-equation (4).

A ferrofluid flow in a magnetic field is accompanied with an
intertwinement of hydrodynamic and magnetic interactions. Being
magnetized, the fluid is subject to magnetic force and
torque with the volume densities $\bf (M\nabla)H$ and $\bf M\times
H$, respectively. On the other hand, the flow vorticity causes a change 
of magnetization. As seen from (1), (4) and (5), $\bf \Omega$ impedes 
alignment of ${\bf M}$ with the direction of the local field ${\bf H}$. 
The appearing magnetic torque is equilibrated by the viscous torque,
\begin{equation}
6\eta \phi ({\bbox{\omega}}-{\bf \Omega })={\bf M\!\times\! H}\,,
\eqnum{6}
\end{equation}
where ${\bbox{\omega}}$ is a macroscopic (i.e., averaged over
physically small ferrofluid volume) angular spin rate of magnetic
grains. But any deviation of ${\bbox{\omega}}$ from ${\bf \Omega}$
leads to an additional dissipation which is just manifested in
rotational viscosity. This dissipation contributes to the stress tensor
[1,2]:
\begin{eqnarray}
\sigma _{ik}=-p\delta _{ik}+\eta \left( \frac{\partial v_{i}}{\partial x_{k}}%
+\frac{\partial v_{k}}{\partial x_{i}}\right) +3\eta\phi\,\epsilon _{ikl}%
(\omega_l-\Omega_l)\nonumber \\
+\frac{1}{4\pi }(H_{i}B_{k}-\frac{1}{2}H^{2}\delta_{ik})\,;
\eqnum{7}
\end{eqnarray}
here $\epsilon _{ikl}$ stands for antisymmetric unit tensor and
the last term represents the Maxwell tensor of magnetic stresses
in ferrofluids. Eliminating ${\bbox{\omega }}-{\bf \Omega }$ from
(7) with the aid of (6), we obtain [1,2]
\begin{eqnarray}
\sigma _{ik}=-\biggl(p+\frac{H^2}{8\pi}\biggr)\delta _{ik}+
\eta \left( \frac{\partial v_{i}}{\partial x_{k}}%
+\frac{\partial v_{k}}{\partial x_{i}}\right)%
+\frac{H_{i}B_{k}}{4\pi}\nonumber\\
+\frac{1}{2}(M_{i}H_{k}-M_{k}H_{i})\,.
\eqnum{8}
\end{eqnarray}
On recognizing that $B_{k}=H_{k}+4\pi M_{k}$, the stress tensor
(8) takes an evidently {\em symmetric} form:
\begin{eqnarray}
\sigma _{ik}=-\biggl(p+\frac{H^2}{8\pi}\biggr)\delta _{ik}+%
\eta \left( \frac{\partial v_{i}}{\partial x_{k}}+\frac{\partial v_{k}}%
{\partial x_{i}}\right)+\frac{H_{i}H_{k}}{4\pi}\nonumber \\%
+\frac{1}{2}(M_{i}H_{k}+M_{k}H_{i})\,, \nonumber
\end{eqnarray}
as it must be in our approximation. Actually, the exact result
reads as $\sigma_{ki}-\sigma_{ik}=I\epsilon_{ikl}(d\omega_l/dt)$
where $I=\rho_sd^2\phi/10$ is the volume density of the particle
moment of inertia ($\rho_s$ the particle density and $d$ their
diameter). However, taking into account an extreme smallness of
$I$ for the particles with $d\sim 10\,{\rm nm}$, we have not
inserted the inertia term $I(d{\bbox \omega}/dt)$ in Eq. (6).

The boundary wall streamlined by the fluid is acted by the force 
$f_i=[\sigma_{ik}]n_k$ on unit area; here $[\;\;]$ denotes
difference evaluated across the fluid--solid interface and $\bf n$
is the normal to the interface. The friction (tangential) force
exerted on the wall is $f_{\tau}=[\sigma_{\tau n}]$. By using the
electrodynamic boundary conditions $[H_{\tau}]=0$ and $[B_n]=0$,
we get from (8)
\begin{equation}
f_{\tau}=\eta(\partial v_{\tau}/\partial x_n)+(M_{\tau}H_n-M_n%
H_{\tau})/2\,. \eqnum{9}
\end{equation}
For the Poiseuille or Couette flow, ${\bf v}=(0,\,v(x),\,0)$, in a
transversal magnetic field, ${\bf H}=(H,\,0,\,0)$, magnetization has 
two components: ${\bf M}=(M_x,\,M_y,\,0)$. Then Eq. (9) can be written 
in the form $f_{\tau}=2(\eta + \eta_r)\Omega $ where 
$2\Omega=\partial v/\partial x $ and rotational viscosity is defined as
\begin{equation}
\eta_r=M_y H/4\Omega\,. \eqnum{10}
\end{equation}
Thus, the additional viscosity is expressed through the off-axis
component of magnetization, $M_y\,$. For small $\Omega\tau$ this
component is also small: according to the all three cited above
magnetization equations, $M_y \propto \Omega\tau$. So, $\eta_r$
does not depend on the flow vorticity in the limit $\Omega\tau\ll%
1$. However, for finite values of $\Omega\tau$ the viscosity does
depend on $\Omega$. As the result, the function $\sigma_{\tau n}%
(\Omega)$ deviates from the linear one, i.e. a ferrofluid acquires
non-Newtonian properties.

Proceeding to the calculation of rotational viscosity on the basis
of Eq. (5), it is convenient to pass from the fields $\bf H$ and
${\bf H}_{\rm e}$ to their non-dimensional values $\bbox \xi $ and
$\bbox \zeta \,$:
\begin{equation}
\frac{d\,{\bbox {\zeta}}}{d\,t}={\bbox {\Omega\times\zeta}}- %
\frac{1}{\tau}\,({\bbox {\zeta}}-{\bbox {\xi}})%
-\frac{L(\zeta)}{2\tau\zeta}\,{\bbox {\zeta}} \times ({\bbox {\zeta}}%
\times {\bbox {\xi }})\,. \eqnum{11}
\end{equation}
At the stated above arrangement of the applied magnetic field with
respect to the fluid flow, Eq. (11) admits a steady solution in
which the effective field ${\bbox \zeta}$ tracks the true field
${\bbox \xi}$ with lag angle $\alpha$, i.e., ${\bbox
\zeta}=(\zeta\cos\alpha,\,\zeta\sin\alpha,\,0)$. The dependence of
$\zeta$ and $\alpha$ upon $\xi$ and $\Omega\tau$ is given by
\begin{equation}
\sqrt{\xi ^2 -\zeta ^2}=\frac{2\Omega\tau\,\zeta}{2+\zeta %
L(\zeta)}\,, \;\;\;\;\;\; \cos\alpha =\frac{\zeta}{\xi} \,.
\eqnum{12}
\end{equation}
Substituting $M_y=nmL(\zeta)\sin\alpha$ in (10) and using (12), we
obtain
\begin{equation}
\eta_r=\frac{3}{2}\eta\phi\,\frac{\zeta L(\zeta)}{2+\zeta %
L(\zeta)}\,.\;\;\;\;\;\;\; {\rm (Sh\,'01)} \eqnum{13}
\end{equation}

By the same way we find from (4)
\begin{equation}
\sqrt{\xi ^2 -\zeta ^2}=\frac{2\Omega\tau\,\zeta L(\zeta)}{\zeta -%
L(\zeta)}\,, \;\;\;\;\;\; \cos\alpha =\frac{\zeta}{\xi} \,,
\eqnum{14}
\end{equation}
that results in
\begin{equation}
\eta_r=\frac{3}{2}\eta\phi\,\frac{\zeta L^2(\zeta)}{\zeta- %
L(\zeta)}\,.\;\;\;\;\;\;\; {\rm (MRSh)} \eqnum{15}
\end{equation}

The solution of Eq. (1) can be presented in the similar form. Let us 
introduce a new variable $\bbox \zeta$ instead of $\bf M$ by the 
relation ${\bf M}=M_0({\bbox \zeta}/\xi) $ where $M_0=nmL(\xi)$. 
It is worth to note that $\zeta$ is no more an effective field unlike 
$\zeta$ in preceding relationships (12)-(15). By substituting the components 
$M_x=M_0(\zeta /\xi)\cos\alpha$ and $M_y=M_0(\zeta /\xi)\sin\alpha$ in (1), 
we get $\cos\alpha=M/M_0=\zeta/\xi$ and
\begin{equation}
\sqrt{\xi ^2 -\zeta ^2}=\frac{2\Omega\tau\,\xi \zeta }{2\xi +%
\zeta ^2L(\xi)}\,. \eqnum{16}
\end{equation}
This and the definition (10) yield
\begin{equation}
\eta_r=\frac{3}{2}\eta\phi\,\frac{\zeta ^2 L(\xi)}{2\xi+ %
\zeta ^2L(\xi)}\,.\;\;\;\;\;\;\; {\rm (Sh\,'72)} \eqnum{17}
\end{equation}

Owing to the smallness of magnetic grains, the Brownian relaxation time 
$\tau$ does not exceed $10^{-4}\,{\rm s}$ even in high-viscous ferrofluids. 
Hence the inequality $\Omega\tau\ll 1$ is usually satisfied. Then, on 
neglecting the value $\Omega\tau$ in (12), (14) and (16), all three these 
relationships are reduced to $\zeta=\xi$. Eliminating now $\zeta$ from (13) 
and (17), we see that both the old (1) and the new (5)
phenomenological equations predict {\em the same} dependence of
rotational viscosity on the magnetic field strength:
\begin{equation}
\eta_{r}(\xi )=\frac{3}{2}\eta \phi \,\frac{\xi L(\xi )}{2+\xi L(\xi )}%
=\frac{3}{2}\eta\phi \,\frac{\xi -\tanh\xi}{\xi +\tanh \xi}\,.\eqnum{18}
\end{equation}
The EFM magnetization equation (4) yields a somewhat different
result. Setting $\zeta=\xi$ in (15) gives
\begin{equation}
\eta _{r}(\xi )=\frac{3}{2}\eta \phi \,\frac{\xi L^{2}(\xi )}{\xi
-L(\xi )} \,. \;\;\;\;\;{\rm (MRSh)} \eqnum{19}
\end{equation}
The viscosities (18) and (19) are compared in Fig. 1. Both of them
approach the saturation value $\eta_r(\infty)=3\eta\phi/2$ at
$\xi\gg 1$. In the figure we plot the reduced rotational viscosity 
$\eta_r(\xi)/\eta_r(\infty)$ as a function of $\xi$. The upper curve 
calculated by the EFM [4] represents a very good approximation. Actually, 
as shown in [5,10], it hardly differs from the exact solution of the 
linearized Fokker-Plank equation. Both the phenomenological equations, 
(1) and (5), result in the lower curve in Fig. 1 that is described by
the Shliomis' formula (18). This function agrees with (19) in the
low- and high-field limits and deviates from that at most on
$15\,\%$ in the entire range of the argument $\xi$.
\begin{figure}
\begin{center}
\epsfxsize=7.8cm
\epsfbox{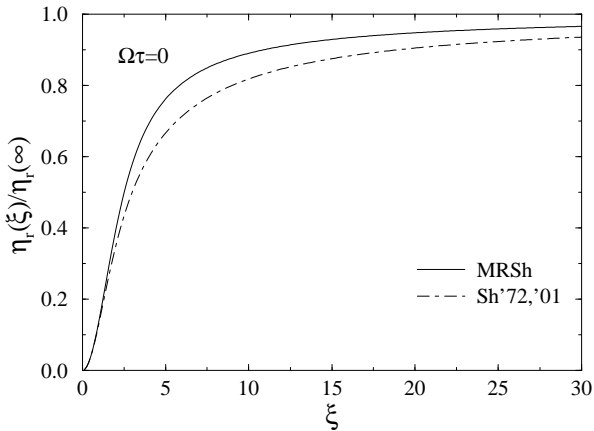}
\end{center}
\vspace*{-0.5cm} \caption{Dependence of the reduced rotational viscosity
on the dimensionless field strength given by (19) (${\rm MRSh}$) and by 
(18) (${\rm Sh}\,'72$ and ${\rm Sh}\,'01$).}
\label{}
\end{figure}
When the ferrofluid is subjected to viscous shear, the magnetic
grains tend to be rotated out of alignment with the magnetic
field. Thus the flow with a sufficiently large shear rate,
$\Omega\tau\geq 1$, induces -- along with the Brownian motion -- a
quotient {\sl demagnetization}. Formally, this effect originates
from decreasing the parameter $\zeta$ determined by Eqs. (12),
(14) and (16). According to these equations, $\zeta=\xi$ when
$\Omega\tau =0$ but the more is $\Omega\tau$ at constant $\xi$, the
less is $\zeta$. The reduction of the magnetization leads in turn to
some decrease in the rotational viscosity. This decrease,
imperceptible in practice up to $\Omega\tau\simeq 1$, becomes then
to be very significant. Figure 2 illustrates the dependence of the
viscosity increase on the magnetic field strength for three values
of the product $\Omega\tau$. Interestingly, under the finite shear
rate the viscosities given by (13) and (17) do not coincide with
each other any more. As seen from the plot, the higher shear the
more discrepancy between viscosity values predicted by the new and
the old phenomenological equations. At high shear in a high field,
Eq. (1) predicts a {\sl hysteresis} of viscosity, which however is
corroborated neither by direct calculations of [6-8] nor by the
solution (14)-(15) of the EFM-equation (4). Our new equation (5)
does also not predict such a hysteresis but it provides us with a
quite satisfactory viscosity description in a wide region of
parameters $\xi$ and $\Omega\tau$. Indeed, in this entire region
the solutions of Eqs. (4) and (5) agree closely, as shown in Fig.
2. Thus, Eq. (5) can be recommended for an employment on a same level
with Eq. (4). It is worth to note that all above calculations,
carried out for a shear flow, apply equally to a rigid rotation of
a ferrofluid with an angular velocity ${\bf \Omega}$ in a constant
transversal magnetic field, ${\bf H\perp\Omega}$, and to a
quiescent ferrofluid subjected to a uniform rotating field ${\bf
H}=(H\cos\Omega t,\,H\sin\Omega t,\,0)$ as well.

The difference between Eqs. (1) and (5) also becomes manifest if
we consider relaxation from an equilibrium magnetization in a
quiescent ferrofluid after the field is suddenly switched off at
the moment $t=0$. Then the fluid remains at rest, ${\bf \Omega}=0$, so 
that $\bf M$ and ${\bf H}_{\rm e}$ in (1) and (5) are always parallel 
to ${\bf H}$. Hence the equations are reduced to
\begin{equation}
\frac{dM}{dt}=-\frac{M-M_0}{\tau}\;\;\;\;\;{\rm and}\;\;\;\;\;
\frac{dH_{\rm e}}{dt}=-\frac{H_{\rm e}-H}{\tau}\,, \eqnum{20{\rm
a,b}}
\end{equation}
respectively. In Fig. 3 we plot the decay of reduced magnetization
$M(t)/M_0$ according to Eqs. (20) with $M_0=nmL(\xi)$ for some
initial field magnitudes $\xi =mH/k_BT$. The solution of Eq.
(20{\rm a}) reads
\begin{equation}
M(t)/M_0=\exp(-t/\tau)\,,\;\;\;\;\;\;\; {\rm (Sh\,'72)} \eqnum{21}
\end{equation}
i.e., it does not depend on $\xi$. Equation (20{\rm b}) has the
solution $H_{\rm e}(t)=H\!\cdot\exp(-t/\tau)$, so that we find
\begin{equation}
M(t)/M_0=L(\xi e^{-t/\tau})/L(\xi)\,.\;\;\;\;\;\;\; {\rm
(Sh\,'01)} \eqnum{22}
\end{equation} 
\begin{figure}
\begin{center}
\epsfxsize=7.8cm
\epsfbox{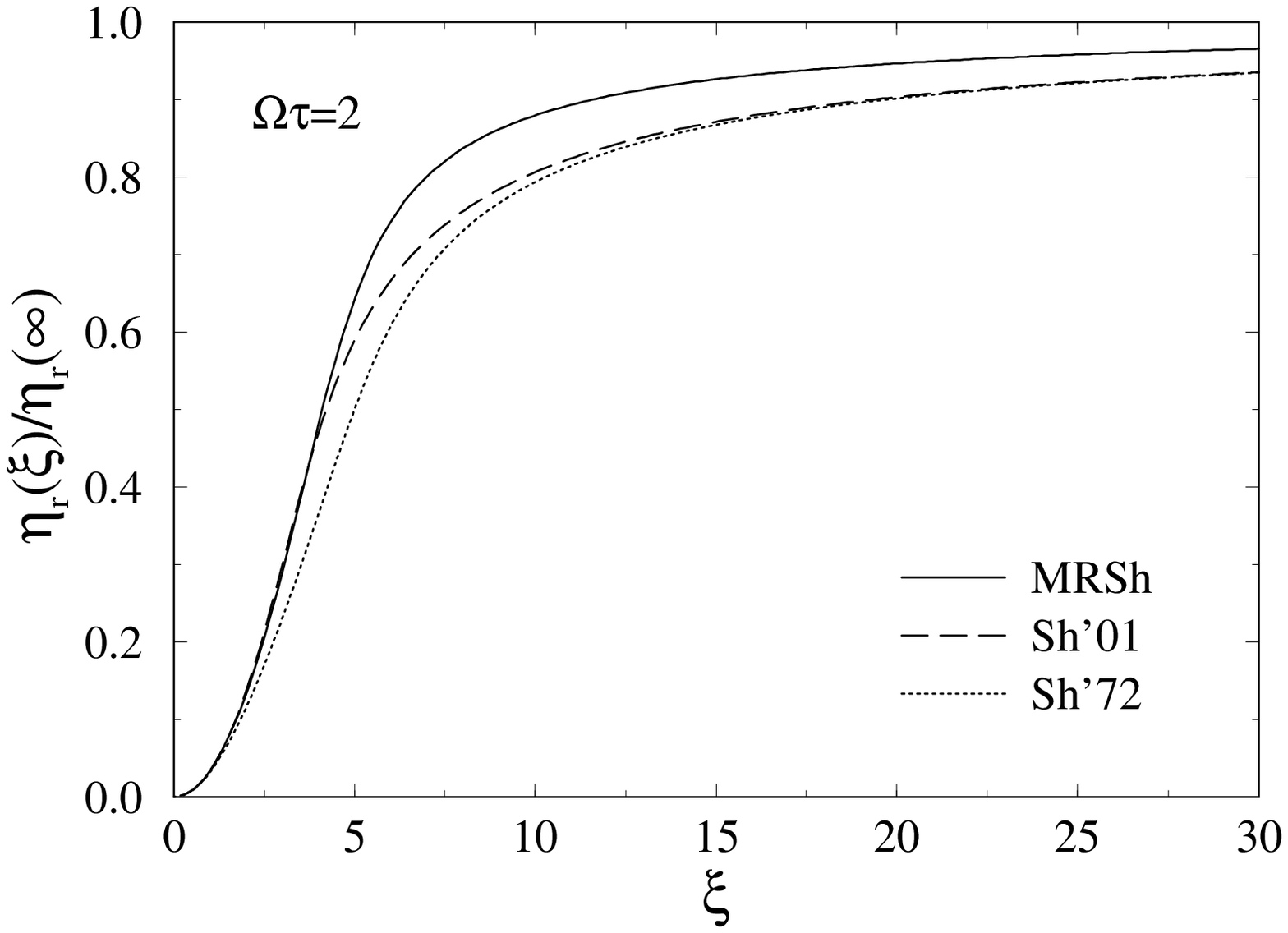}
\epsfxsize=7.8cm
\epsfbox{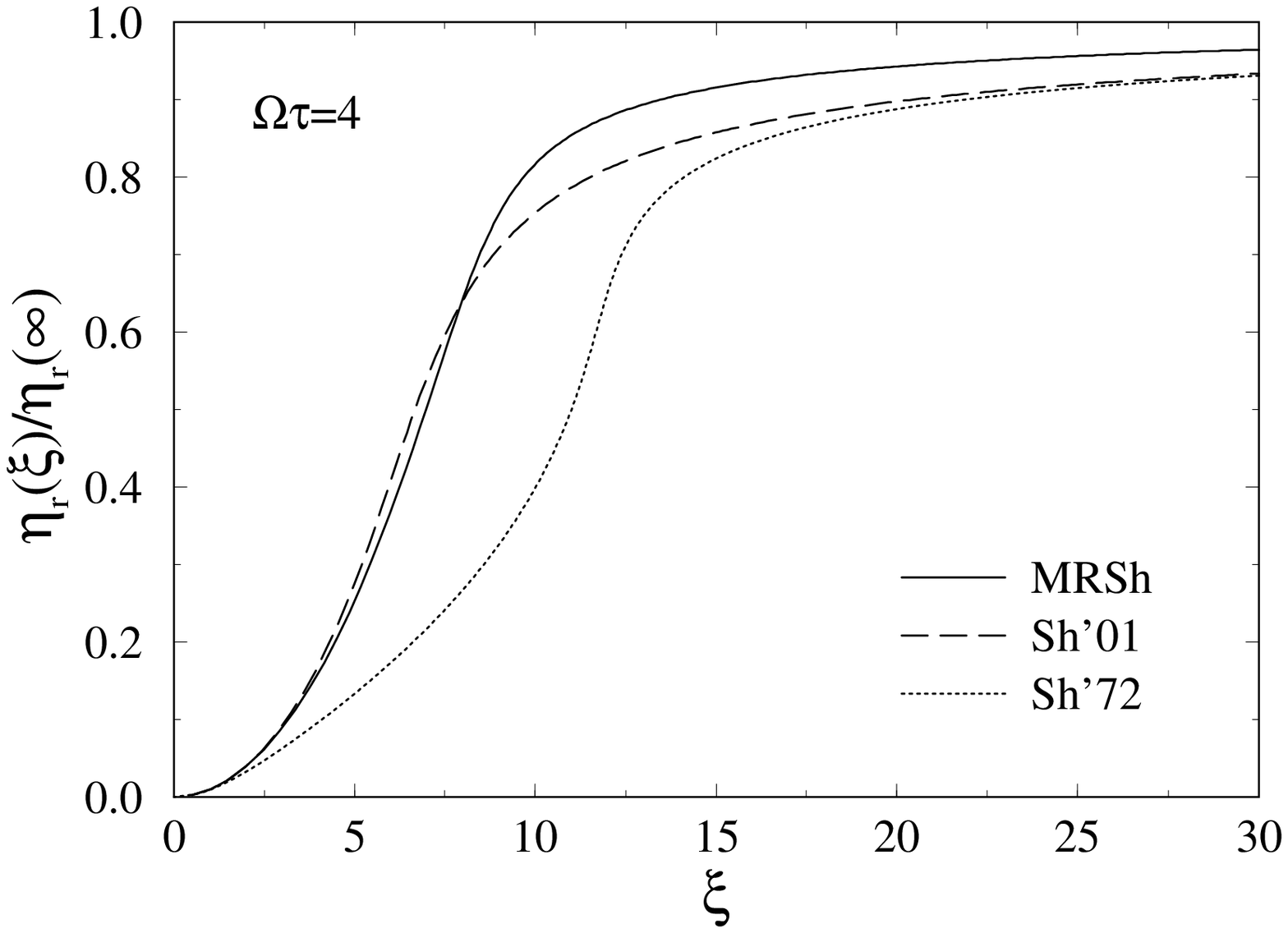}
\epsfxsize=7.8cm
\epsfbox{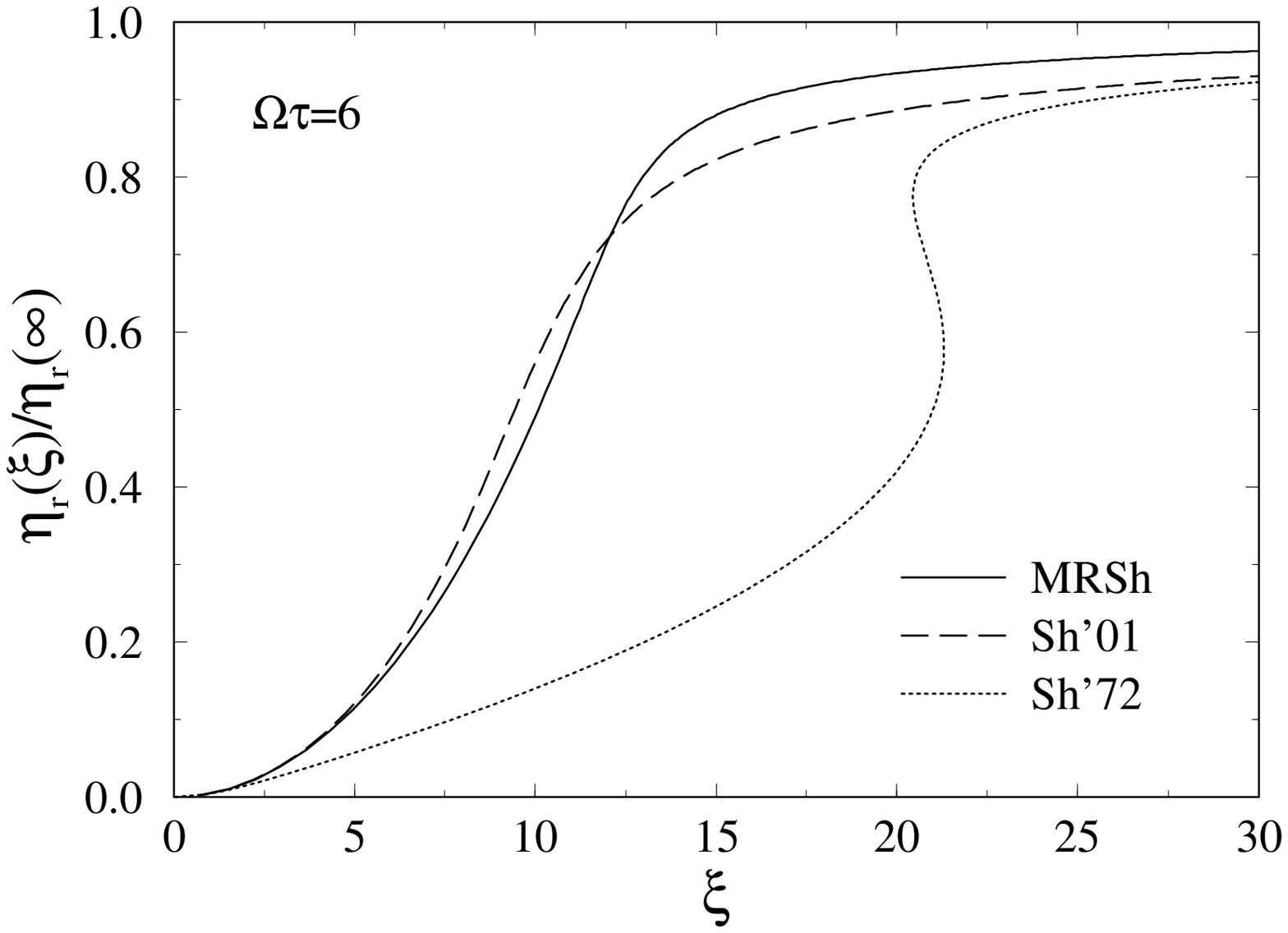}
\end{center}
\vspace*{-0.5cm} \caption{Dependence of the rotational viscosity on the field 
for some values of the shear rate $\Omega\tau$, as calculated from the EFM
[(14)-(15), ${\rm MRSh}$], and by the new [(12)-(13), ${\rm Sh}\,'01$] and 
old [(16)-(17), ${\rm Sh}\,'72$] phenomenological approaches.}
\label{}
\end{figure}

Thus, only in the limit $\xi\ll 1$ both decays, (21) and (22),
are exponential and coincide with each other. Because ferrofluids
exhibit nonlinear magnetization, the decay (22) is {\sl
nonexponential} for a finite $\xi$ magnitude. Side by side with
the dependence of viscosity on magnetic field and vorticity, this
distinct difference in relaxation behavior may be of relevance for
the interpretation of experiments and testing the magnetization
equation (5).

\begin{figure}
\begin{center}
\epsfxsize=7.8cm
\epsfbox{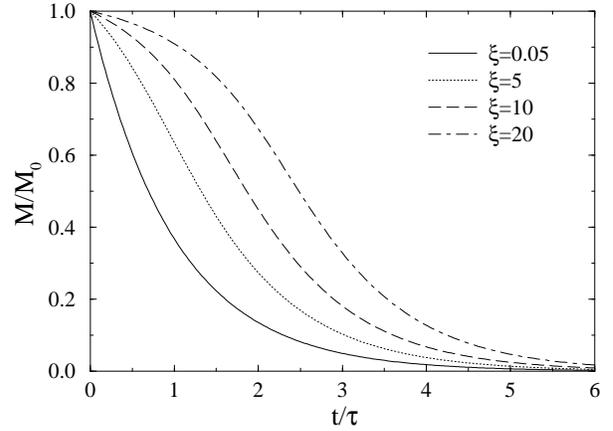}
\end{center}
\vspace*{-0.5cm} \caption{Time dependence of the reduced magnetization 
$M(t)/M_0$ after the
magnetic field $\xi$ is switched off, as described by (22). The lowest 
curve represents also the solution (21) for any $\xi$ values.}
\label{}
\end{figure}

\vskip 0.1 truecm

This work was supported by the Meitner--Humboldt Research Award
adjudged by the Alexander von Humboldt Foundation.

% For wide abstract
\end{multicols}

\end{document}